\documentstyle[twocolumn,pra,aps]{revtex}

\begin{document}

\bibliographystyle{prsty}

\draft

\tighten

\title{Decoherence in localized photon emission}

\author{Holger F. Hofmann}
\address{Institut f\"ur Technische Physik, DLR\\
Pfaffenwaldring 38--40, D--70569 Stuttgart, Germany}

\maketitle

\section{Introduction}
The conventional method of describing optical emissions from quantum systems 
treats this process as a transition between non-localized eigenstates of the 
Hamiltonian. This represents a lack of resolution in time and space. While 
the advances in short time spectroscopy have already removed much of the
limitation in temporal resolution, this must yet be achieved for the spatial 
resolution. However, advances in mesoscopic physics should eventually make 
such observations possible, opening the door to new phenomena of quantum 
coherence.

The calculations presented in sections 3 and 4 are based on a simple picture 
of near field spectroscopy, in which the local channels are only sensitive to 
emissions from one of two interacting two level systems. The nonlocal farfield
modes are also included. The probability of photon detection in the localized 
channels is determined as a function of time and emission sequence, which 
mirrors the spatial evolution of the excitation in the observed system.

\section{Time resolved photon emission}
Our model for photon emission is based on a spatiotemporal interpretation of 
Wigner-Weisskopf theory \cite{Hof95}. If we describe the transition from an 
excited state $\mid e \rangle$ to the ground state $\mid g \rangle$, with an 
energy difference of $\hbar\omega_0$ between these two states of the system, 
and choose the wavevektor k to describe the states of the photon after 
emission, with $kc=\omega$ as the dispersion of the evenly spaced field 
states near the resonance, then the wave function $\mid \psi(t) \rangle$
for the emission process is described by the amplitudes
\begin{mathletters}
\begin{eqnarray}
\langle e; vac.\mid \psi(t) \rangle &=& e^{-i \omega_0 t - \Gamma t/2} \\
\langle g; k \mid \psi(t) \rangle &=&
\sqrt{\frac{c\Gamma}{2\pi}} e^{-ikct}\frac{1-e^{-\Gamma t/2+ikc-i\omega_0 t}}
{kc-\omega_0+i\Gamma/2}.
\end{eqnarray}
\end{mathletters}

This rather unintuitive result can be Fourier transformed to a one dimensional
real space representation. In the case of the three dimensional light field, 
the one dimensional coordinate r represents the radius, while the angular 
dependence is defined by the symmetry of the transition, which is usually 
dipole radiation. For a more complete discussion of these questions, see 
\cite{Hof95}. The amplitudes of the single photon states now read
\begin{eqnarray} 
\langle g; r \mid \psi(t) \rangle &=& -i\sqrt{\frac{\Gamma}{c}} 
e^{(\Gamma/2+i\omega_0)(x/c-t)} \nonumber \\
&&  \mbox{for}\; 0<x<ct, \;\mbox{else}\; 0
\end{eqnarray}
A decisive advantage of the real space representation is that it reveals the 
physical assumptions behind the approximations of Wigner Weisskopf theory. 
While the temporal evolution of the k-space result is difficult to interpret 
since it seems to include complicated reabsorption processes, the real space 
representation shows that the photon wavefunction moves with the constant 
velocity c away from the system, retaining the phase which the system had at 
the time of emission. There is no reabsorption in this base and the propagator
of the field is a translation with constant velocity c, without dispersion, 
so it is possible to assign to each real space state at r an emission time 
t-r/c.This model of photon emission therefore provides a simple approach to 
questions of time resolved spectroscopy.

\section{The system}
We now apply this model to emissions from a system of two identical two state 
systems, which may exchange their excitations via a Forster interaction. The 
Hamiltonian of this system is given by
\begin{eqnarray}
\hat{H}&=& 2 E_0 \mid ee \rangle\langle ee\mid  
          +  E_0 \mid eg \rangle\langle eg\mid 
          +  E_0 \mid ge \rangle\langle ge\mid \nonumber \\ 
      &&  +  W   \mid ge \rangle\langle eg\mid 
          +  W   \mid eg \rangle\langle ge\mid 
\end{eqnarray}
where the base states have been chosen as the product states of the excited 
states e and the ground states g of the two identical two state systems. 
The Eigenstates of this Hamiltonian are given by 
\begin{eqnarray}
\mid 2(+)\rangle &=& \mid ee \rangle \nonumber \\
\mid 1(+)\rangle &=& \frac{1}{\sqrt{2}}(\mid eg \rangle + \mid ge \rangle) 
\nonumber \\
\mid 1(-)\rangle &=& \frac{1}{\sqrt{2}}(\mid eg \rangle - \mid ge \rangle) 
\nonumber \\
\mid 0(+)\rangle &=& \mid gg \rangle 
\end{eqnarray}
separated into symmetrical $(+)$ and antisymmetrical $(-)$ states with respect
to an exchange of the two identical systems. The Hamiltonian in this base is
\begin{eqnarray}
\hat{H} &=& 2 E_0 \mid 2(+) \rangle\langle 2(+)\mid \nonumber \\  
        && + (E_0+W) \mid 1(+) \rangle\langle 1(+)\mid \nonumber \\
        && + (E_0-W) \mid 1(-) \rangle\langle 1(-)\mid
\end{eqnarray}

The next step is to define the modes of the light field coupled to the system.
In the conventional situation, when no optical nanostructures are present and 
the size of the system is far less than the wavelength, only the nonlocal far 
field mode N couples to the system. However, in order to achieve spatial 
resolution, local modes which couple to the two identical parts of our system 
separately are necessary as well. In order to maintain the symmetry, we 
introduce the two local modes L1 and L2. L1 couples exclusively to transitions
in the first system (from $\mid ee \rangle$  to $\mid ge \rangle$  and from 
$\mid eg \rangle$ to $\mid gg \rangle$), while L2 couples in the same manner 
to transitions in the second system (from $\mid ee \rangle$  to $\mid eg 
\rangle$  and from $\mid ge \rangle$ to $\mid gg \rangle$).

The base states of the light field in real space are therefore 
$\mid r, N(+)\rangle$,$\mid r, L1\rangle$ and $\mid r, L2\rangle$. To 
reflect the symmetry of the system, it is convenient to transform the local 
fields to their symmetrical and antisymmetrical base states:
\begin{eqnarray}
\mid r, L(+)\rangle &=& 
\frac{1}{\sqrt{2}}(\mid r, L1 \rangle + \mid r, L2\rangle) \nonumber \\
\mid r, L(-)\rangle &=& 
\frac{1}{\sqrt{2}}(\mid r, L1 \rangle - \mid r, L2\rangle) 
\end{eqnarray}
Since the unitary evolution of the system will preserve the parity, a 
transition between an antisymmetrical state and a symmetrical state requires 
the emission of an antisymmetrical photon, whereas all transitions within the 
symmetric system emit only symmetrical photons. Due to this property of the 
system, it is now very easy to determine the temporal evolution of the 
emission processes of the system.

\section{Temporal evolution and measurements}
We are now ready to describe the emission process, starting from any excited 
state of the system. The most interesting case is the emission of two photons 
from the $\mid ee\rangle$ state, since the measurement of the two time 
correlation of the consecutive emissions should provide information on the 
coherent evolution of the system. To simplify our equations, however, it is 
sufficient to start from the states with a single excitation, since the 
measurement of the first photon emitted at time $t=0$ effectively projects 
the system into a well defined state with a single excitation left. In the 
following we investigate the dynamics after the emission of a photon in L2 
at $t=0$. This is equivalent to the dynamics of $\mid \psi (t=0)\rangle
= \mid eg \rangle$. The probabilities of emission calculated from this 
initial state represent the conditional probability of finding emissions in 
the L1, L2 and N channels, given that a photon was detected in L2 at $t-r/c=0$.
With $\Gamma_N$ and $\Gamma_L$ as the emission rates of the nonlocal and 
local modes respectively, and defining $\omega_+ = (E_0+W)/\hbar$ and 
$\omega_- = (E_0-W)/\hbar$ , the evolution of the wavefunction is
\begin{mathletters}
\begin{eqnarray}
\langle 1(+);vac.&\mid& \psi(t)\rangle =
\frac{1}{\sqrt{2}} e^{-(\Gamma_L+\Gamma_N)t/2-i\omega_+ t}\\
\langle 1(-);vac.&\mid& \psi(t)\rangle =
\frac{1}{\sqrt{2}} e^{-\Gamma_L t/2-i\omega_- t}\\
\langle 0(+);r,N(+)&\mid& \psi(t)\rangle = \nonumber \\
-i\sqrt{\frac{\Gamma_N}{2c}} && e^{-(\Gamma_L/2+\Gamma_N/2+i\omega_+)(r/c-t)}\\
\langle 0(+);r,L(+)&\mid& \psi(t)\rangle = \nonumber \\
-i\sqrt{\frac{\Gamma_L}{2c}} && e^{-(\Gamma_L/2+\Gamma_N/2+i\omega_+)(r/c-t)}\\
\langle 0(+);r,L(-)&\mid& \psi(t)\rangle = \nonumber \\
+i\sqrt{\frac{\Gamma_L}{2c}} && e^{-(\Gamma_L/2+i\omega_-)(r/c-t)}
\end{eqnarray}
\end{mathletters}
With these amplitudes, we can calculate all measurement probabilities. The 
effects of local coherence should be visible in the probabilities of 
measureing a photon in the L1 or the L2 channels. We therefore calculate the 
conditional probability of measuring a photon in L1 at a time t after the 
measurement of a photon in L2, assuming that both measurements occur at an 
equal distance r from the system:
\begin{eqnarray}
p_{L2,L1}(t) &=& 
    \mid \langle 0(+);r,L1\mid \psi(t+r/c)\rangle \mid^2 \nonumber \\
&=& 1/2 (\mid \langle 0(+);r,L(+)\mid \psi(t+r/c)\rangle \mid^2
    \nonumber \\
& & + \mid \langle 0(+);r,L(-)\mid \psi(t+r/c)\rangle \mid^2) \nonumber \\
& & + Re (\langle 0(+);r,L(+)\mid \psi(t+r/c)\rangle \nonumber \\
& & \hspace{0.5cm}\langle \psi(t+r/c) \mid 0(+);r,L(-)\rangle) \nonumber \\
&=& \frac{\Gamma_L}{4c} e^{-\Gamma_L t}
[1+2 e^{-\Gamma_N t/2}\cos(\delta\omega t) + e^{-\Gamma_N t}] 
\end{eqnarray}
with $\delta\omega = 1/2 (\omega_+ - \omega_-) = W/\hbar$. The resulting time 
dependence describes the two time correlation between the emission events. 
Figure 1 illustrates this correlation for a typical choice of 
parameters, $\Gamma_L \ll \Gamma_N$ and $\delta\omega = 8 \Gamma_N$. 
In general, there are three timescales independently given by 
$\Gamma_L$, $\Gamma_N$ and $\delta\omega$. Since nano-optical modes
can be expected to have a much weaker coupling to the system than the far 
field modes, the assumptio that $\Gamma_L \ll \Gamma_N$ is quite realistic, 
while the choice of $\delta\omega$ remains somewhat arbitrary. The figure 
clearly shows the quantum beats of the excitation, which oscillates between 
the two systems at a frequency of $\delta\omega$. These coherent beats decay 
on a timescale of $\Gamma_N$. The quantitative expression for this decay is
the ratio of the amplitude $A$ and the mean value $M$ of the beats:
\begin{equation}
\frac{A}{M} = \frac{2 e^{\Gamma_N t/2}}{1+e^{\Gamma_N t}} 
= \frac{1}{\cosh (\Gamma_N t/2)}
\end{equation} 
Therefore, the effect of the far field on the observation of localized 
quantum beats is to cause decoherence. The decoherence is strongly 
non-exponential for $\Gamma_N t < 1$.

\section{Interpretation}
The temporal evolution of the wave function clearly shows that the fast 
transitions from $\mid 1(+)\rangle$ to  $\mid 0(+)\rangle$ quickly reduce the 
amplitude of $\mid 1(+)\rangle$ relative to $\mid 1(-)\rangle$. Therefore the 
coherence between $\mid 1(+)\rangle$ and $\mid 1(-)\rangle$ is lost. This can 
be interpreted in terms of measurement theory as a potential measurement 
of $\mid 1(+)\rangle$ in the N channel, which necessarily disturbs the 
distinction between $\mid eg \rangle$ and  $\mid ge \rangle$. The differenc
in the emission rates of $\mid 1(+)\rangle$ and $\mid 1(-)\rangle$ 
means, that a photon detected in a local channel is most likely to be emitted 
by the $\mid 1(-)\rangle$ state, if $\Gamma_N t \gg 1$.

The presence of channels coupling both locally and nonlocally can be 
interpreted as a continuous weak measurement of non-commuting observables 
\cite{Hel74}. The decoherence in the local channels is then seen to arise 
from the quantum fluctuations of the field in the nonlocal channel. Note that 
the effect of the nonlocal channel is therefore seen even in those 
experiments, during which no photon is emitted into this channel (e.g. the 
L2-L1 emission discussed above). This possibility to observe the presence of 
the nonlocal mode even without the excitation of a photon in this channel 
is similar in nature to the possibility of an interaction free measurement 
discussed by Elitzur and Vaidmann in \cite{Eli93}.

\section{Conclusions}
The model presented here allows the description of simultaneous coupling 
to local and nonlocal modes by formulating the complete wave function of 
the system and the interacting fields. The results clearly show how 
quantumbeats occuring between the two local subsystems decohere due to 
the influence of the nonlocal channel. Therefore, this model is a useful 
tool in the investigation of the relationship between coherent and 
dissipative phenomena observed with high spatial and temporal resolution.

The Author would like to thank G. Mahler for many helpful discussions.


%
%
\begin{figure}
\caption{Two time correlation of the L2-L1 emission sequence.}
\end{figure}
%

\end{document}